\documentclass[aps,prb,twocolumn,notitlepage,showpacs,floatfix,superscriptaddress,10pt,
groupedaddress,superscriptaddress]{revtex4-1}
\usepackage{times,amsmath,amsfonts,amssymb,mathrsfs,graphics,graphicx,color,comment,bm}
\usepackage[next]{inputenc}
\usepackage[dvips]{epsfig}
\usepackage[pdfstartview=FitH]{hyperref}
\usepackage{natbib}
\usepackage{pdfsync}
\usepackage{appendix}
\def\be{\begin{equation}}
\def\ee{\end{equation}}

\def\bi{\begin{itemize}}
\def\ei{\end{itemize}}
\def\bn{\begin{enumerate}}
\def\en{\end{enumerate}}
\def\bea{\begin{eqnarray}}
\def\eea{\end{eqnarray}}

\def\ba{\begin{array}}
\def\ea{\end{array}}
\def\bd{\begin{displaymath}}
\def\ed{\end{displaymath}}

\def\bra#1{{\langle #1 \vert}}
\def\ket#1{{\vert #1 \rangle}}

\begin{document}

\title{Ground-state fidelity of the spin-1 Heisenberg  
chain with single ion anisotropy: quantum renormalization group and 
exact diagonalization approaches
}

\author{A. Langari}
\affiliation{Department of Physics, Sharif University of Technology, P.O.Box
11155-9161, Tehran, Iran}
\affiliation{Center of excellence in Complex Systems and Condensed Matter
(CSCM), Sharif
University of Technology, Tehran 1458889694, Iran}
\affiliation{Max-Planck-Institut f\"ur Physik komplexer Systeme, 01187 Dresden,
Germany}
\email{langari@sharif.edu}
\homepage{http://sharif.edu/~langari/}
\author{F. Pollmann}
\affiliation{Max-Planck-Institut f\"ur Physik komplexer Systeme, 01187 Dresden, Germany}
\author{M. Siahatgar}
\affiliation{Max-Planck-Institut f\"ur Physik komplexer Systeme, 01187 Dresden, Germany}

\newcommand{\frank}[1]{ { \color{red} \footnotesize (\textsf{FP})
\textsf{\textsl{#1}} } }

\newcommand{\abdollah}[1]{ { \color{blue} \footnotesize (\textsf{AL})
\textsf{\textsl{#1}} } }






\date{\today}

\begin{abstract}
We study the phase diagram of the anisotropic spin-1 Heisenberg chain with 
single ion anisotropy (D) using a ground-state fidelity approach.
The ground-state fidelity and its corresponding susceptibility are 
calculated within the quantum renormalization group scheme where we obtained
the renormalization of fidelity preventing to calculate the ground state.
Using this approach, the  phase boundaries between the antiferromagnetic
N\'{e}el, Haldane and large-$D$ phases are obtained for the whole phase
diagram, which justifies the application of quantum renormalization group 
to trace the symmetery protected topological phases.
In addition, we present numerical exact diagonalization (Lanczos) results 
in, which we employ a recently introduced non-local order parameter to locate the
transition from Haldane to large-$D$ phase accurately.
\end{abstract}


\maketitle

\section{Introduction\label{introduction}}

A major challenge in the physics of correlated electrons is to understand
quantum phases and the  transitions between different phases. Quantum spin
chains have proven to be extremely useful model systems as these relatively
simple models can exhibit very complex phase diagrams.\cite{Sachdev:2001} The
different phases are usually characterized by spontaneous symmetry breaking
and can be understood in the framework of Landau's theory of phase transition.
\cite{Landau37} Based on the idea of symmetry breaking, local order parameters can be
defined that distinguish the different phases. Over the past few decades,
several phases have been discovered , which do not break any symmetry and thus
cannot be understood using local order parameters. Such phases are known as
``topological phases''. A well known  example of a topological phase is the
Haldane phase in spin-chains with odd integer spin.\cite{Haldane-1983a,Haldane-1983}
The Haldane phase  is an example of a so-called \emph{symmetry protected
topological phase} (SPTP). In a series of works, it has been shown that
SPTP's can be completely characterized using cohomology theory.
\cite{Chen-2011,Chen-2011a,Pollmann:2012,Pollmann:2010,Schuch-2011} 
In the case of
the Haldane phase any of the following symmetries is sufficient to protect
the phase: the dihedral group of $\pi$-rotations about two orthogonal axes,
time-reversal symmetry, or bond centered inversion symmetry.

Transitions between different quantum phases are accompanied by a qualitative change of the ground state, driven
by quantum fluctuations as a control parameter in the Hamiltonian is tuned.
Apart from few exactly solvable models (e.g., the transverse field Ising model), the precise determination of the quantum critical
point is a very challenging task -- even if a local order parameter changes at the transition.  If no order parameter exists to distinguish the phases, the task is  even more difficult. Generically, phase transitions are characterized by detecting a non-analytic behavior in
some physical properties of the system. Intensive investigations have been devoted to characterize quantum phase transition using quantum information concepts.  \cite{Amico:2008,Amico:2009}
It turned out that the ground-state fidelity is a particular powerful quantity to identify a quantum phase transition, irrespective of the existence of global symmetry breaking. Thus it is very useful when studying topological phase transitions.\cite{Zanardi:2006, Zhou:2008b}
An essential change of the ground state at the quantum critical point leads to an abrupt drop of ground
state fidelity accompanied by a divergent peak of its corresponding susceptibility.\cite{Gu:2010}

Recently it has been proposed to use the  ``quantum renormalization group'' (QRG) method to obtain
the quantum-information related properties of strongly correlated systems.  \cite{Kargarian:2007,Kargarian:2008,Kargarian:2009}
The QRG  is a technique based on coarse graining to the most
important degrees of freedom in the low energy spectrum.\cite{Wilson:1975,Pfeuty:1982} It has been shown that the ground-state fidelity can be obtained in terms of QRG procedure without the actual need to obtain the ground state. \cite{Langari:2012,Amiri:2013}
Motivated by this approach to investigate quantum critical behaviors in terms of QRG for ground-state fidelity, it is very interesting to apply this technique to topological quantum phase transitions.

In this paper, we consider  the anisotropic spin-1 Heisenberg chain with
single ion anisotropy. Besides various symmetry broken phases, this model has
two phases, which do not break any symmetry: The Haldane phase, i.e., a SPTP
is found around the isotropic point, and in the presence of a strong ion
anisotropy the system is in the so-called large-$D$ phase (which is
adiabatically connected to a simple site factorized state). We apply the QRG
algorithm to calculate the ground-state fidelity and its corresponding
susceptibility. Moreover, we present  exact diagonalization (Lanczos)
results on finite clusters , which are compared to the QRG results. We
calculate non-local order parameters based on the inversion symmetry (which
protects the Haldane phase) to obtain a precise determination of quantum
critical point on finite clusters. Here we use the fact, that the expectation
value of any member of permutation operator on a closed ring is negative in
the Haldane phase while it becomes positive in the large-$D$ phase.\cite{Pollmann:2012}

The remainder of this paper is organized as follows: In the next section we introduce the model and summarize some known results. In Sec.~\ref{c} we explain how to use the QRG approach
to obtain the ground-state fidelity and analyze its behavior close to quantum critical points, which leads to the phase diagram of the spin-1 chain.
We present  exact diagonalization results in Sec. \ref{ed} where we also introduce the parity order
parameter, which changes sign at the Gaussian critical point. Finally, we present a summary and we conclude our results in Sec. \ref{summary}.

\section{Model}
\begin{figure}[tb!]
  \begin{center}
    \includegraphics[width=75mm]{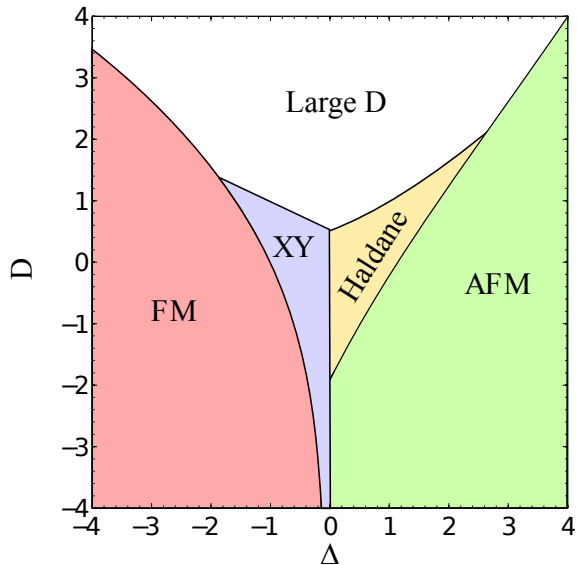}
    \caption{(color online) A sketch of the spin-1 phase diagram based on  results from Ref.~\onlinecite{Chen:2003}.}
    \label{S1pd}
  \end{center}
\end{figure}

We focus on the anisotropic spin-1 Heisenberg chain with a single ion anisotropy ($D$). The model Hamiltonian is given by
\begin{equation}
 H=J\sum_{i=1}^N \big[S_i^x S_{i+1}^x+S_i^y S_{i+1}^y+\Delta S_i^z S_{i+1}^z
+D(S_i^z)^2\big],
\label{Hamiltonian}
\end{equation}
where $J$ is the antiferromagnetic exchange coupling, $\Delta$ is the easy axis anisotropy
and $D$ represents the single-ion anisotropy parameter. The phase diagram of
Hamiltonian Eq.~(\ref{Hamiltonian}) is well established\cite
{Botet,Schulz:1986,Kitazawa96,Chen:2003,Tzeng:2008,CamposVenuti:2007,Hu:2011}
and sketched in Fig.~\ref{S1pd}.  It contains a gapless XY,  a ferromagnetic
(FM) and an anti ferromagnetic (AF) phase. The increase of $D$ drives the
ground state of model from Haldane to large-$D$ phase through a Gaussian
transition.\cite{Schulz:1986} The lack of a local order parameter for the
Gaussian transition and its topological nature make the study very
challenging. The phase diagram has been studied using a density matrix
renormalization group (DMRG) computation of ground-state fidelity for both the
isotropic \cite{Tzeng:2008b} and anisotropic \cite{Tzeng:2008} chain.  It was found for the
isotropic model ($\Delta=1$) that  the fidelity susceptibility diverges at the phase
transitions out of the Haldane phase according to scaling relations defined
in Ref.~\onlinecite{CamposVenuti:2007}. However, the fidelity susceptibility fails to
 detect the Gaussian transition for an anisotropy parameter of $\Delta=0.5$. The scaling analysis
performed in Ref.~\onlinecite{Tzeng:2008} found a critical exponent for the
correlation length of $\nu\simeq 1.51$. Extensive DMRG simulations \cite
{Hu:2011}, in which the length of chain goes to $N=10000$  and the number of
kept states can reaches $m=1000$, were implemented to study the quantum critical
properties of spin-1 chain accurately. However, the exponent reported in the latter work (Ref.~\onlinecite{Hu:2011})
is $\nu\simeq 2.387 (5)$ for $\Delta=0.5$. This discrepancy represents the non-local nature of strong correlation
in the ground state of model specially close to quantum critical point, which makes the precise determination
of quantum critical properties a challenging task.

\section{Quantum renormalization group approach and ground-state fidelity \label{c}}

The QRG proceeds by keeping the most important degrees of freedom
while integrating out the rest within an iterative procedure. As a result, the original Hamiltonian
is mapped into a renormalized Hamiltonian defined by the set of renormalized couplings.
The renormalization of couplings generates the flow of couplings, which describes the quantum phase diagram
in addition to the ground state properties of the model.

\begin{figure}
 \includegraphics*[width=\linewidth]{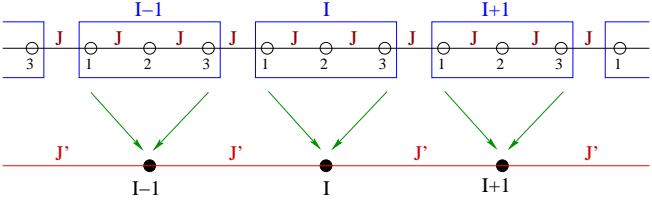}
\caption{(color online) Decomposition of the chain into blocks. Each block will be represented
by an effective spin (filled-circle) after the renormalization with renormalized
interactions.}
\label{blocks}
\end{figure}

The standard QRG prescription, which is implemented in this work, is based on Kaddanoff block
renormalization group.\cite{Pfeuty:1982,Delgado:1996,Sierra:1997,Langari:1998,Jafari:2006,Jafari:2007}
The algorithm of this procedure can be expressed in the following steps:
(i) The lattice is decomposed into isolated blocks (see Fig.~\ref{blocks}) where the Hamiltonian is written as 
a sum of isolated block Hamiltonian ($H^B$) and inter-block interactions ($H^{BB}$), i.e., $H=H^B+H^{BB}$.
(ii) The block Hamiltonian is diagonalized exactly and some of the low-lying
energy eigenstates of each block are kept to build up an embedding (projection)
operator ($T$), representing the most important subspace of the original Hilbert space ($\mathcal{H}$).
(iii) The original Hamiltonian ($H$) is mapped into the renormalized (effective) Hamiltonian ($H'$) utilizing
the embedding operator, i.e.,
\be
H'=T^{\dagger} H T.
\label{hprime}
\ee
The renormalized Hamiltonian, which is supposed to be self similar
to the original one, defines the renormalization of coupling constants (QRG-flow).

In our implementation of the QRG procedure, the model Eq.~(\ref{Hamiltonian})  is decomposed
into individual blocks of three spins, see Fig.~\ref{blocks}, where the intra-block and inter-block Hamiltonians
are given by the following relations
\bea
\label{hb-hbb}
&&\hspace{1cm} H^B=\sum_{I=1}^{N/3} h_I^B, \hspace{1cm} H^{BB}=\sum_{I=1}^{N/3} h_{I,I+1}^{BB}, \nonumber \\
&&h_I^B=J\Big[\sum_{j=1}^2 \big(S_{I,j}^x S_{I,j+1}^x+ S_{I,j}^y S_{I,j+1}^y+\Delta S_{I,j}^z S_{I,j+1}^z \big) \nonumber \\
&& \hspace{2cm} +D \sum_{j=1}^3 (S_{I,j}^z)^2 \Big], \nonumber \\
&&h_{I,I+1}^{BB}=J\big(S_{I,3}^x S_{I+1,1}^x+ S_{I,3}^y S_{I+1,1}^y+\Delta S_{I,3}^z S_{I+1,1}^z\big).
\eea
Here $S_{I,j}^{\alpha}$ denotes the $\alpha$-component of the $j$-th spin in block $I$.
The energy eigenstates of $h_I^B$ are calculated exactly and the three lowest eigenvectors are
denoted by $\ket{\phi_0}$ and $\ket{\phi_{\pm}}$ with the corresponding eigenvalues $E_0$ and $E_1$, respectively.
The first excited state energy ($E_1$) is doubly degenerate (corresponding to $S^z=\pm1$)
Accordingly, the embedding operator for each block is constructed by
\be
T_I=\ket{\phi_+}\bra{+1}+\ket{\phi_0}\bra{0}+\ket{\phi_-}\bra{-1},
\label{embedding}
\ee
where $\ket{\pm1}, \ket{0}$ are
the base kets for the renormalized Hilbert space of each block.
The projection of the original Hamiltonian to the renormalized Hilbert space of the whole system
is done by the global embedding operator ($T=\otimes_{I=1}^{N/3} T_I$)
as defined in Eq.~(\ref{hprime}).
The renormalized Hamiltonian ($H'$) is akin to the original one, Eq.~(\ref{Hamiltonian}),
replacing the couplings with the following renormalized ones,
i.e.,
\be
J'\equiv J'(J, D, \Delta)\;,\; D'\equiv D'(D, \Delta) \;,\; \Delta'\equiv \Delta'(D, \Delta) .
\label{rg-flow}
\ee
The explicit form of renormalized couplings in terms of the original ones
and some details of the renormalization procedure are presented in Appendix~\ref{qrg-appendix}.
The ground state properties of the model can be extracted from the QRG-flow or equivalently from the effective
operators in the renormalized Hilbert space ($\mathcal{H'}$).
The renormalized Hamiltonian ($H'$) defined above ensures that if $\ket{\psi'_0}$ is the ground state of $H'$,
$\ket{\psi_0}=T \ket{\psi'_0}$ is the ground state of $H$ imposing $T^{\dagger} T=\mathbf{1}_{\mathcal{H'}}$.

Next we discuss a prescription to calculate the ground-state fidelity within the QRG approach.\cite{Langari:2012} 
In this respect, we recall the definition of the ground-state fidelity $F$, for a finite system size $N$,
\be
F\equiv F(\lambda, \delta, N)= \langle {\psi_0(\lambda_{-})}\ket{\psi_0(\lambda_{+})},
\label{fidelity}
\ee
where $\ket{\psi_0(\lambda_{\pm})}$ is the normalized ground state at arbitrary coupling
$\lambda_{\pm}\equiv \lambda \pm \delta/2$ and $\delta$ stands for a small variation of $\lambda$.
According to the QRG prescription, the fidelity can be expressed in terms of the ground state of
the renormalized Hamiltonian utilizing the embedding operator,
\bea
F&=&\bra{\psi'_0(\lambda_{-})} T^{\dagger}(\lambda_{-}) T(\lambda_{+}) \ket{\psi'_0(\lambda_{+})} \nonumber \\
&=&\bra{\psi_0(\lambda'_{-})} T^{\dagger}(\lambda_{-}) T(\lambda_{+}) \ket{\psi_0(\lambda'_{+})}.
\label{renormalized-fideltiy}
\eea
In fact the group property of renormalization justifies that
$\ket{\psi'_0(\lambda)}=\ket{\psi_0(\lambda')}$ where $\lambda'$ is the renormalized coupling.
The operator product $T^{\dagger}(\lambda_{-}) T(\lambda_{+})$ establishes
the renormalization of fidelity, which provides $F$ in terms of the fidelity of the
renormalized ground state ($F'$). In the underlying model, the coupling $\lambda$ is composed of two
parameters $\Delta$ and $D$, which requires to define a small deviation for both couplings,
namely $\delta_{\Delta}$ and $\delta_{D}$. A straightforward calculation leads to the following
expression
\be
T^{\dagger}_I(D_{-}, \Delta_{-}) T_I(D_{+}, \Delta_{+})=
\omega_I \mathbf{1}_I +\upsilon_I (S_I^z)^2,
\label{t-dagger-t}
\ee
where $\omega_I$ and $\upsilon_I$ are scalar functions of the coupling constants
($D_{-}, \Delta_{-}; D_{+}, \Delta_{+}$).
Moreover, the ground states of all blocks, which are taken into account in the calculation of
Eq.~(\ref{renormalized-fideltiy}) belong to $S^z=0$ sector, which renders zero value for the second term
in Eq.~(\ref{t-dagger-t}). Therefore, including the contribution of all blocks the renormalization of
fidelity is given by
\be
F=
\Big(\omega_I(D_{-}, \Delta_{-}; D_{+}, \Delta_{+}) \Big)^{N/3} F'
\label{fidelity-renormalization}
\ee
where $F'$ is the renormalized fidelity.
The procedure is iterated $m$-times associated with the size of system $N=3^{m+1}$, which connects the
original fidelity ($F$) to the fidelity of a system, which has been renormalized $m$-times ($F^{(m)}$),
\be
F=\Big(\prod_{n=0}^{m-1} \big[\omega_I(D^{(n)}_{-}, \Delta^{(n)}_{-};
D^{(n)}_{+}, \Delta^{(n)}_{+})\big]^{\frac{N}{3^{n+1}}}\Big) F^{(m)}.
\label{iterative-fidelity}
\ee

\begin{figure}
 \includegraphics*[width=\linewidth]{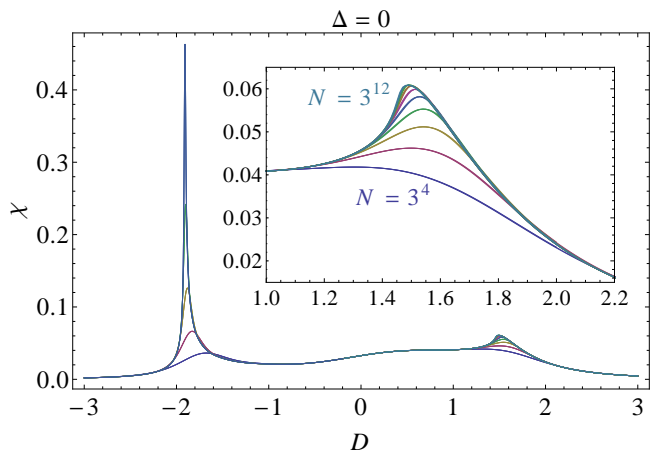}
\caption{(color online) \label{Chi-QRG-Delta-0}
Fidelity susceptibility ($\chi$) versus $D$ at $\Delta=0$ for different sizes
$N=3^{m+1}, m=3, \dots, 11$ and $\delta=0.001, 0.005, 0.01$.
The size dependence is only visible around the critical points presented by different colors.
The left peak corresponds to the N\'{e}el-Haldane quantum critical point while the right one
presents the Haldane to large-$D$ quantum phase transition.
Inset: The right peak around $D\simeq 1.5$ is plotted in larger scale to clarify the behavior
at the Haldane--large-$D$ quantum critical point.
}
\end{figure}

For simplicity we calculate the ground-state fidelity at fixed $\Delta$ and two slightly different
values of $D$, namely $D\pm\delta_D/2$ for three different values of $\delta_D \equiv \delta=0.001, 0.005, 0.01$.
Hence, we track the quantum critical points by adjusting $D$.
An abrupt drop of fidelity in the vicinity of quantum critical point is a consequence of an essential change in the
structure of ground state, which is usually accompanied by a divergence/maximum of fidelity susceptibility.
Therefore, fidelity or its corresponding susceptibility are reliable signatures of quantum criticality.
As far as the ground state
is normalized to unity, the leading term of the fidelity is expressed by
\be
F(D, \delta; \Delta, N)\backsimeq 1- \frac{N \delta^2}{2} \chi(D; N),
\label{susceptibility}
\ee
where $\chi$ is the fidelity susceptibility, which is defined by
$\chi \equiv \frac{-1}{N}\frac{\partial^2 F}{ \partial \delta^2}$.

\begin{figure}
 \includegraphics*[width=\linewidth]{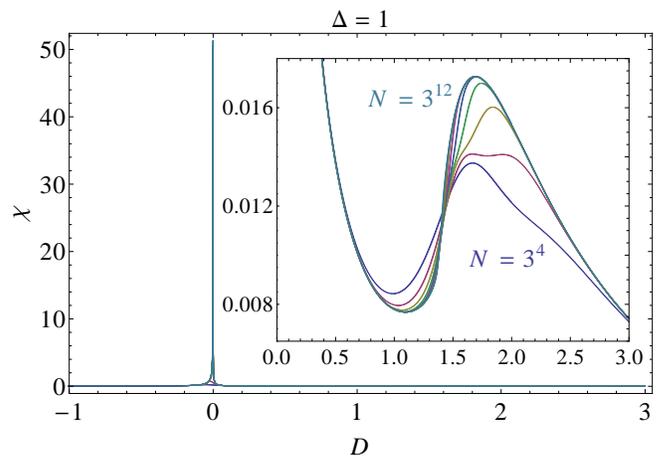}
\caption{(color online) \label{Chi-QRG-Delta-1}
Fidelity susceptibility ($\chi$) versus $D$ at the isotropic point ($\Delta=1$) for different sizes
$N=3^{m+1}, m=3, \dots, 11$ and $\delta=0.001, 0.005, 0.01$.
The very sharp peak at $D\simeq 0$, which is associated to the N\'{e}el-Haldane quantum critical point
induces a scale, which interdicts another peak at $D\sim 1.7$.
Inset: $\chi$ vs. $D$ for $0 < D <3$ in larger scale such that the peak on Haldane--large-$D$ phase
transition is clearly observable.
}
\end{figure}

We have plotted $\chi$ versus $D$ at $\Delta=0$  in Fig.~\ref{Chi-QRG-Delta-0} for different values of $\delta$
and system sizes. Accordingly, $\chi$
extinguishes the dependence on $\delta$ such that all plots with different $\delta=0.01, 0.005, 0.001$ fall on each other
except at the critical points where the size dependence turns out. Close to a quantum critical
point $\chi$ exhibits a peak, which intensifies by increasing the system size. However, for  off-critical
regions there is no size dependence on $\chi$, which is clearly visible from Fig.~\ref{Chi-QRG-Delta-0}.
The left peak in Fig.~\ref{Chi-QRG-Delta-0} represents the N\'{e}el to Haldane phase transition where
N\'{e}el order exists for $D\lesssim-2.0$ while the right peak corresponds to the Haldane--large-$D$ quantum
phase transition. The right peak around $D\sim 1.45$ is plotted in larger scale as an inset in Fig.~\ref{Chi-QRG-Delta-0}
to justify its size dependence as a signature of quantum phase transition.
Similarly for the isotropic case ($\Delta=1$), the fidelity susceptibility versus $D$ is plotted
in  Fig.\ref{Chi-QRG-Delta-1} for the same values of $\delta=0.01, 0.005, 0.001$ and system sizes
$N=3^{n}, n=4, \dots, 12$. Two signatures of quantum phase transition are observed, the left sharp
peak is clearly seen while the second peak appears on a smaller scale around $D\sim 1.7$.
To find out the second peak we have plotted the region $0 < D <3$ in larger scale as an inset in
Fig.~\ref{Chi-QRG-Delta-1}, which verifies an enhancing peak around $D\sim 1.7$ that
is corresponding to the Haldane--large-$D$ phase transition.

In both Figs.~\ref{Chi-QRG-Delta-0} and \ref{Chi-QRG-Delta-1} the left peak in $\chi$ is higher
and sharper than the right one. The reason why the two transition show such a drastically different behavior is not completely clear. One might speculate that the reason is inherited to the two different mechanism for
the N\'{e}el-Haldane and Haldane--large-$D$ phase transitions. The former is associated with a
spontaneous symmetry breaking, which sets up a nonzero staggered magnetization in the N\'{e}el phase
while the latter corresponds to a 
quantum phase transition to a symmetry protected topological phase. However, the QRG approach to fidelity is able to detect both types of phase transition as
witnessed above. 


\begin{figure}
 \includegraphics*[width=\linewidth]{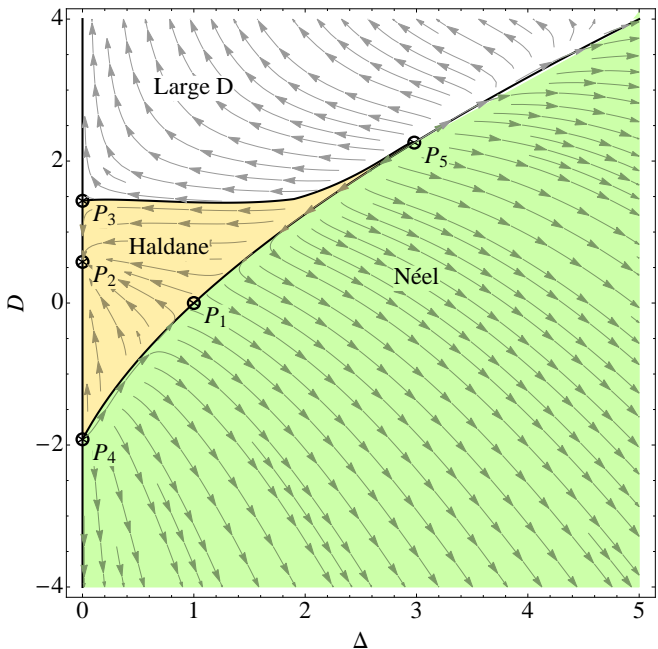}
\caption{(color online) The QRG-flow, which specifies the phase diagram of $S=1$ XXZ model with single
ion anisotropy (D) where $\Delta$ stands for easy-axis anisotropy coupling. The critical boundaries are
denoted by black lines and $\circledast$ represent the fixed points
P$_1$ - P$_5$.
The phase diagram is composed of N\'{e}el, Haldane and large-$D$ phases. }
\label{phase-diagram}
\end{figure}

As described above, the QRG fidelity approach can be used to find the quantum phase boundary
of our model. Moreover, the analysis of the QRG-flow Eq.~(\ref{rg-flow})
 gives a clear picture of the topography of
ground state phase diagram as
depicted in Fig.~\ref{phase-diagram}.
The QRG-flow has two types of fixed points, those which are attractive
in all directions stand for stable phases and those which are repulsive at least in one direction
represent  quantum critical points.
The fixed points ($D^*, \Delta^*$) are denoted by $\circledast$ in Fig.~\ref{phase-diagram} and are
labeled  P$_1$:(0.0, 1.0), P$_2$:(0.58, 0.0),
P$_3$:(1.45, 0.0), P$_4$:(-2.0, 0.0), P$_5$:(2.5, 3.0) and two others
for extremely large couplings, namely ($\infty$, 0.0) and $(-\infty, \infty)$.
P$_1$ is associated to the isotropic $S=1$ Heisenberg model (without single ion anisotropy)
which retains the SU(2) symmetry upon QRG transformation and belongs to the Haldane phase.
The SU(2) symmetric fixed point (P$_1$)
is unstable upon adding axial anisotropy either by
easy-axis or single-ion types. The fixed point P$_1$ is isotropic, i.e., it has the full SU(2) symmetry,
while P$_2$ has a $U(1) \times Z2$ symmetry.
The black line, which passes through P$_1$ specifies the phase boundary between the N\'{e}el
and Haldane phases. All couplings in the closed bounded region labeled ``Haldane'' run to the
stable fixed point P$_2$ under QRG transformations while P$_3$ is
unstable in vertical direction and corresponds to the critical point
between Haldane and large-$D$ phases on the $\Delta=0$ line. The black line
which originates from P$_5$ and ends at P$_3$ represents the border between
Haldane and large-$D$ phases as a function of $\Delta$. On the negative part of vertical axis
the unstable fixed point P$_4$ stands for the critical point between N\'{e}el and Haldane phases.
The tri-critical point P$_5$ is associated to the point where the two borders
between N\'{e}el-Haldane and Haldane--large-$D$ merge and the model belongs to either
N\'{e}el or large-$D$ phases.  

We showed that using the QRG approach to the  ground-state fidelity, we can obtain the 
right topology of the phase diagram. However,  the phase boundaries are  
shifted compared to the known (numerically exact) results. \cite{Chen:2003}

\section{Exact diagonalization \label{ed}}

\begin{figure}
\includegraphics*[width=\linewidth]{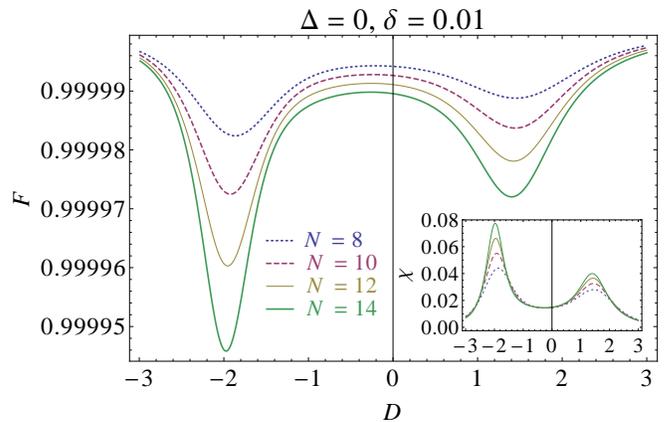}
\caption{(color online) \label{Fidelity-ED-Delta-0}
Exact diagonalization results of fidelity (F) versus $D$ for $\Delta=0$, $\delta=0.01$ and
on chain lengths $N=8, 10, 12, 14$. The corresponding fidelity susceptibility ($\chi$) is plotted
as an inset.
}
\end{figure}

In the following, the ground-state fidelity of the Hamiltonian defined in Eq.~(\ref{Hamiltonian})
is calculated exactly for finite sizes using the Lanczos method. Here, we compute the fidelity for two ground states with slightly different values of $D$ at fixed $\Delta$,
i.e.
\be
F=\langle \psi_0(\Delta, D-\delta/2) \ket{\psi_0(\Delta, D+\delta/2)},
\ee
where $\delta=0.01$. We have plotted fidelity
versus $D$ in Fig.~\ref{Fidelity-ED-Delta-0} for $\Delta=0$ and even number of
sites on periodic chain $N=8, 10, 12, 14$. The fidelity has two minima
at $D\sim -2$ and $D\sim 1.5$ where clear finite size effects are observed, which are signatures
of quantum phase transition. Moreover, we plot the fidelity susceptibility ($\chi$) versus $D$
as  inset in Fig.~\ref{Fidelity-ED-Delta-0} where two peaks associated to the minima of fidelity
confirm the existence of the quantum critical points. The peak of susceptibility has an apparent
maximum where its corresponding height ($\chi_m$) scales like $\chi_m \sim N^{\alpha}$.
For the inset of Fig.~\ref{Fidelity-ED-Delta-0} the left peak scales linearly with system size ($N$)
and the right peak, which corresponds to the Haldane-large-$D$ phase transition scales as $N^{0.63}$.
A similar analysis for $\Delta=1$ shows that the left peak of $\chi$ around $D\simeq 0$ scales
with $N^{0.807}$ and for the right one corresponding to Haldane-large-$D$ transition $\chi_m\sim N^{0.561}$.
In all cases the fidelity susceptibility diverges in the thermodynamic limit although its divergence
is linear only for the  N\'{e}el-Haldane phase transition at $\Delta=0$ and is sub-linear for
other cases specially for the Haldane to large-$D$ phase transition.

\begin{figure}
\includegraphics*[width=\linewidth]{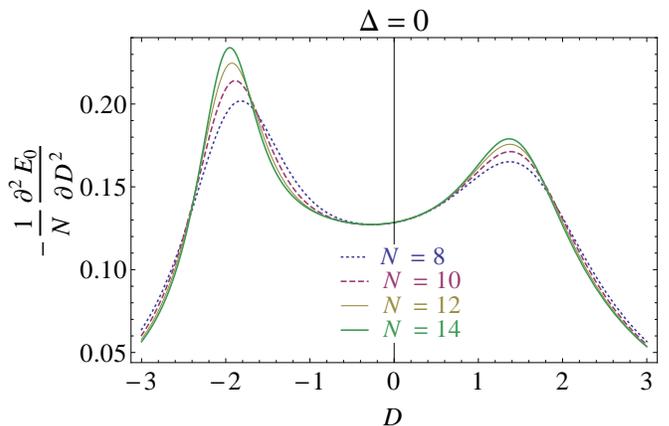}
\caption{(color online) \label{minus2ndGSEnergy-ED-Delta-0}
Second derivative of ground state energy density versus $D$ for $\Delta=0$.
}
\end{figure}

According to Ref.~\onlinecite{CamposVenuti:2007}, a quantum phase transition does not always
lead to a superextensive growth of fidelity susceptibility
which means that $\chi$ can be even finite as $N\rightarrow \infty$ for gapless systems.
However, our analysis on small system sizes ($N\leq14$) shows that the fidelity susceptibility is always divergent
both on the N\'{e}el-Haldane and Haldane-large-$D$ phase transitions. Accordingly, we would expect
to observe similar peaks in the 2nd derivative of ground state energy density.
We have plotted  $\frac{-1}{N}\frac{\partial^2 E_0}{\partial D^2}$ versus $D$ for $\Delta=0$
in Fig. \ref{minus2ndGSEnergy-ED-Delta-0} where $E_0$ is the ground state energy of a chain of $N$ spins
within periodic boundary condition. As is expected the peaks in susceptibility (inset of Fig.~\ref{Fidelity-ED-Delta-0})
are sharper than the corresponding ones in the 2nd derivative of energy close to quantum critical
points manifesting that $\chi \sim \frac{1}{E^2_{gap}}$ while
$\frac{\partial^2 E_0}{\partial D^2} \sim \frac{1}{E_{gap}}$ where $E_{gap}$ is the energy gap.
Although the results of exact diagonalization shows a divergent behavior of $\chi$ and clear size
dependent maximum on the 2nd derivative of energy density for $\Delta=0$, the position of
peak $D_m$ for different sizes does not satisfy the scaling relation
$|D_m(N)-D_c| \sim N^{-1/\nu}$ where $D_c$ is the quantum critical value and $\nu$ is the exponent
which shows the divergence of correlation length.\cite{CamposVenuti:2007} Therefore, the
accurate determination of the quantum critical point should be obtained from the specific property of
the model, which will be discussed in next subsection.

\begin{figure}
\includegraphics*[width=\linewidth]{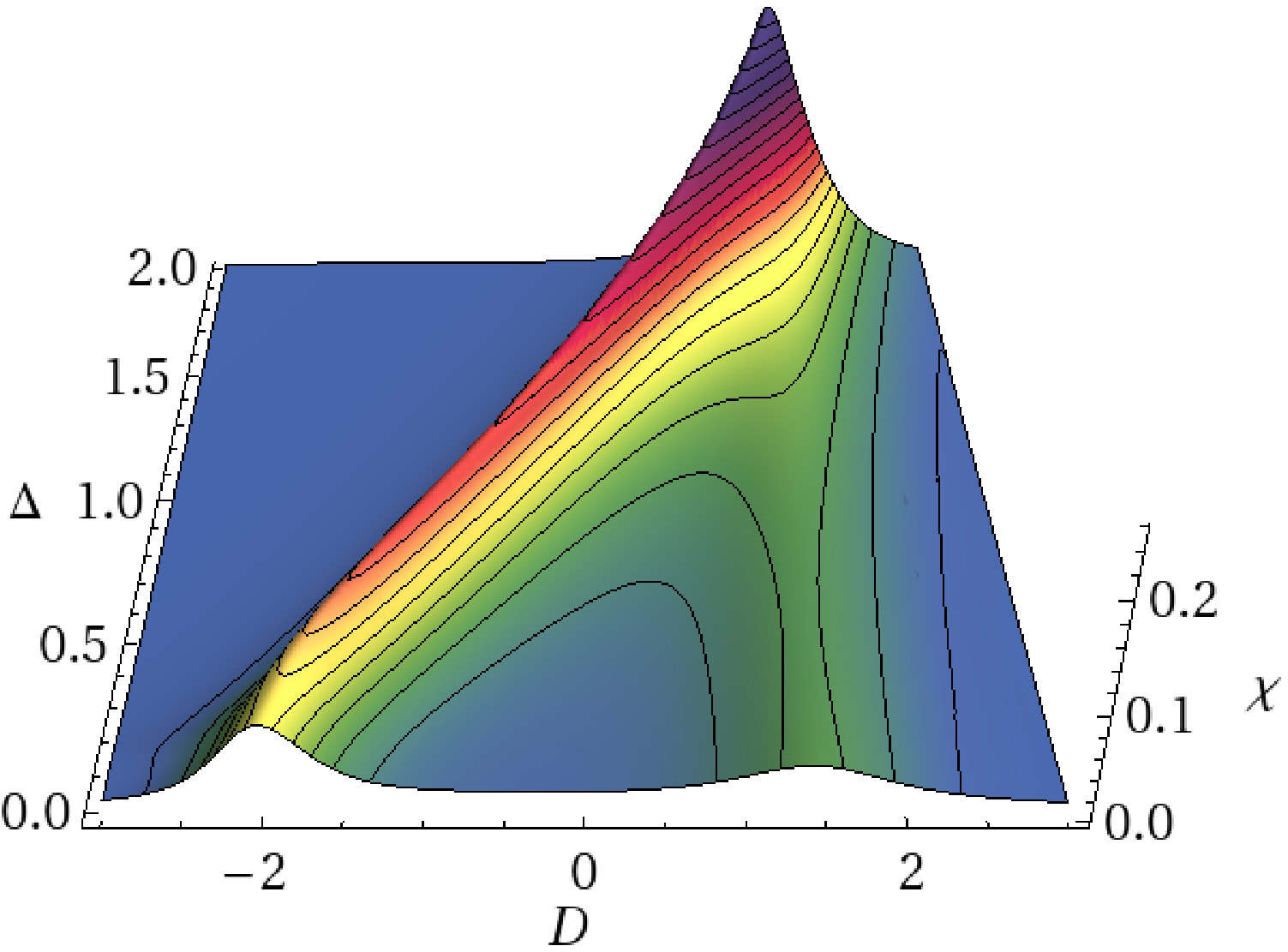}
\includegraphics*[width=0.6\linewidth]{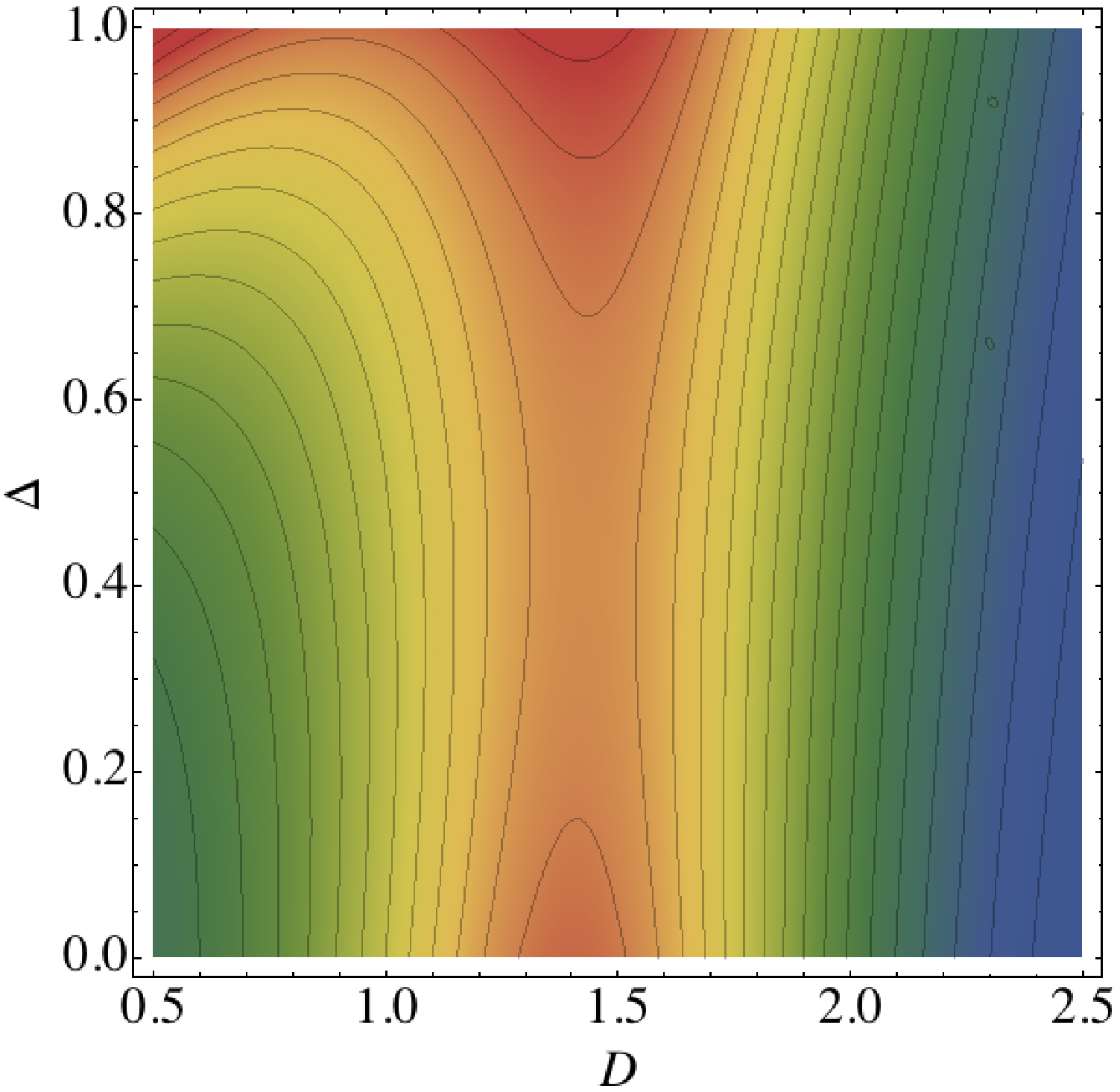}
  \caption{(color online) \label{Chi-ED-3DPlot}
Top: The 3D-plot of fidelity susceptibility ($\chi$) versus $D-\Delta$ plane.
The peaks of $\chi$ correspond to the critical phase boundaries
where the higher peaks represent the border between N\'{e}el-Haldane phases
and the lower peaks is a representative for the Haldane to large-$D$ quantum
phase transition.
Bottom: The contour plot of Haldane to large-$D$ transition in larger scale for
more clear vision.}
\end{figure}

To get more insight on the fidelity behavior of the whole phase diagram we have computed the
fidelity for all parameters in the range of $-3 \leq D \leq +3$ and
$0 \leq \Delta \leq 2$. The exact diagonalization is performed on a periodic chain of
$N=14$ and the fidelity is calculated at fixed $\Delta$ for two ground states with
slightly different $D$ values, namely $D-\delta/2$ and $D+\delta/2$ with $\delta=0.01$.
We present the three-dimensional plot of fidelity susceptibility versus $D-\Delta$ plane
in Fig.~\ref{Chi-ED-3DPlot}-(top). The grid points on the $\Delta$ axis is also $0.01$.
A clear set of peaks similar to the inset of Fig.~\ref{Fidelity-ED-Delta-0} is observed
in Fig.~\ref{Chi-ED-3DPlot}-(top)
which is the signature of quantum phase transition. The left set of peaks, which is stronger than
the right one is a representative of N\'{e}el to Haldane phase transition while the right
set of peaks corresponds to the Haldane to large-$D$ phase boundary. We  also
show the contour plot of the  peaks in  Fig.\ref{Chi-ED-3DPlot}-(bottom)
at different scale to show the boundary more clearly. The three dimensional plot of $\chi$ is
consistent with the quantum phase diagram of Fig.\ref{phase-diagram} confirming that ground-state fidelity
is a good indicator to find quantum phase transition.

\subsection{Parity order parameter}
Here we discuss how the bond inversion symmetry of the Hamiltonian can be used to determine the phase transition from 
the Haldane phase to the large-$D$ phase very accurately from exact diagonalization studies. 
In Refs.~\onlinecite{Pollmann:2010,Pollmann:2012} it was shown that the $S=1$ Haldane phase 
is protected by the link inversion symmetry. A simple picture helps to realize the nature of the 
two phases: Let us suppose that the Haldane phase is described
by the valence bond solid (VBS) configuration where each spin $S=1$ is composed of two $S=1/2$ spins 
and the ground state is a sequence of singlets formed between two neighboring $S=1/2$ spins. 
We now consider an odd number of $S=1$ spins on a periodic chain and invert the lattice about its reflection 
symmetric plane (see Fig.~2 of Ref.~\onlinecite{Pollmann:2012}). The ground state of the Haldane phase gets 
a $\pi$-phase under inversion as
a result of inverting an odd number of singlet bonds. However, the large-$D$ phase is totally symmetric under 
reflection, which returns a phase of $0$. Therefore the ground states are in different symmetry sectors in the 
two phases and a level crossing occurs at the critical point (i.e., the ground state in the Haldane phase is 
in the sector, which is odd under inversion and the large-$D$ phase in the sector, which is even). 
This distinction has been used in exact diagonalization of $S=2$
chains to detect symmetry protected phases.\cite{Tonegawa:2011} In rings with an even number of sites, 
however, the ground state is always in the even sector and thus we do not expect a level crossing. 
In both cases, it is still possible to distinguish the phases by calculating the expectation value of 
an operator, which inverts a block of consecutive spins.\cite{Pollmann:2012b} This order parameter can 
been seen as a generalization of the string order parameter introduced in Ref.~\onlinecite{Nijs:1989}.

For the implementation of the inversion symmetry based order parameter, we define permutation operators $\mathcal{P}_{i,j}$, which exchange the positions of two spins $i$ and $j$:
\bea
\mathcal{P}_{i,j} = \vec{S}_i \cdot \vec{S}_j &+& (\vec{S}_i \cdot \vec{S}_j)^2 - \mathbf{1},
\label{permutation}
\eea
where $\vec{S}_i$ is the spin-1 operator at site $i$ and $\mathbf{1}$ is
the identity operator.

\begin{figure}
\includegraphics*[width=\linewidth]{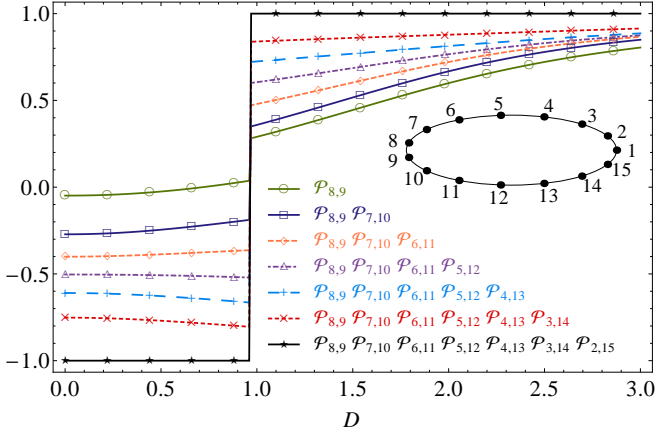}
\caption{(color online) \label{Inversion-N-Odd}
The ground state expectation value of different permutation operators
versus $D$ on $N=15$ (odd-N) isotropic spin $S=1$
chain with periodic boundary conditions. The Haldane phase returns
negative value while it jumps to a positive value for the large-$D$ phase.
}
\end{figure}

We begin by considering the case of a ring with an odd number of spins. The expectation value of different permutations of the $N=15$ ring versus $D$ is shown in Fig.~\ref{Inversion-N-Odd}.
The expectation value is always negative in the Haldane phase and jumps to a positive value upon arriving
in the large-$D$ phase, clearly distinguishing the two phases. The reflection symmetry, which is expressed by a two-point permutation, $P_r=\mathcal{P}_{2,15}\mathcal{P}_{3,14}\mathcal{P}_{4,13} \mathcal{P}_{5,12}\mathcal{P}_{6,11}\mathcal{P}_{7,10}\mathcal{P}_{8,9}$ returns $-1$ in the Haldane phase and $+1$ for the large-$D$ one. 
We have calculated $\langle \psi_0| P_{r} |\psi_0 \rangle$ on
periodic chain with $N=9, 11, 13, 15$ and found the value of $D_p$ at which
it changes sign form -1 to +1. The finite size analysis of our data by extrapolating
with $D_p=D_c+ a N^{-b}$ gives the critical point ($D_c$) in the thermodynamic limit.
We have presented the critical point between the Haldane and large-$D$ phases
within 3-digits of accuracy in Table~\ref{t1}. A comparison with recent large scale DMRG results \cite{Hu:2011} for $\Delta=0.5$ and $1$ shows a
perfect agreement up to the 3rd digit. This is remarkable as our results are obtained 
from  exact diagonalization on small clusters, which require much less resources.
\begin{table}
\caption{(color online) \label{t1}
The critical point ($D_c$) between the Haldane and large-$D$ phase
for different values of anisotropies ($\Delta$).}
 \begin{tabular}{|c||c|c|c|c|c|c|c|c|}
\hline
 $\Delta$  & 0 & 0.1 & 0.2 & 0.3 & 0.4 & 0.5 & 0.6 & 0.7 \\ \hline
 $D_c$ & 0.347 & 0.403 & 0.458 & 0.515 & 0.575 & 0.636 & 0.698 & 0.763 \\
\hline \hline
 $\Delta$  & 0.8 & 0.9 & 1 & 1.1 & 1.2 & 1.3 & 1.4 & 1.5 \\ \hline
 $D_c$ & 0.830 & 0.898 & 0.968 & 1.039 & 1.113 & 1.187 & 1.265 & 1.343 \\
\hline
\end{tabular}
\end{table}

\begin{figure}
\includegraphics*[width=\linewidth]{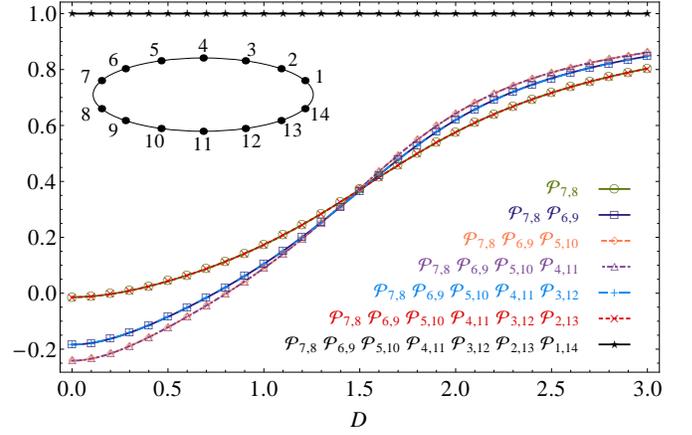}
\caption{(color online) \label{Inversion-N-Even}
The ground state expectation value of different permutation operators
versus $D$ on N=14 (even-N) isotropic spin $S=1$
chain with periodic boundary conditions. The Haldane phase returns
negative value while it gets positive continuously for the large-$D$ phase.
}
\end{figure}

Next we examine the ground state expectation value of the permutation
operators on an even-N chains with periodic boundary condition. We have plotted the
permutation expectation value of N=14 versus $D$ in Fig. \ref{Inversion-N-Even}. 
The expectation value still changes its sign when going from the Haldane phase into the large D phase, however, finite size effects are much stronger in this case and an accurate determination of the critical point is not possible. As expected, the order parameter does not show a jump and we obtain always $+1$ when inverting the full chain, i.e., the ground state is always in the even sector.

\section{Summary and conclusions \label{summary}}

We have studied the anisotropic spin-1 Heisenberg chain with single ion anisotropy
by utilizing quantum renormalization group and exact diagonalization methods.
We have applied the recent implementation of QRG to calculate the ground-state fidelity \cite{Langari:2012}
of the spin-1 chain without the need to calculate the ground state exactly.
We have obtained the renormalization of fidelity as an analytic expression in terms of the
renormalized coupling constants for the whole phase diagram. The fidelity
shows a drop at the quantum critical point irrespective of being due to global
symmetry breaking like N\'{e}el-Haldane transition or a symmetry protected topological
phase transition like Haldane-large-$D$ transition. It confirms that fidelity is
an appropriate candidate to discriminate the quantum phase transitions even if it is a topological one.
We have also obtained the fidelity susceptibility, which shows a growing peak at the quantum
critical point by increasing size of system where its qualitative behavior does not depend on the easy axis anisotropy.
The growing peak is divergent at the N\'{e}el-Haldane critical point while it is just a maximum
at the Haldane-large-$D$ phase transition. Although the quantitative results of QRG fidelity is not
accurate, its qualitative description of the phase diagram is fairly good in addition to the RG-flow
which gives a topography of the phase diagram.

We have also analyzed the fidelity and its susceptibility data, which come from
exact diagonalization simulation on finite periodic chains ($N=8, \dots,15$). Two very different types of behaviors
have been observed for even and odd number of sites. For even number of sites, $N=8, 10, 12, 14$, the
fidelity shows a drop at the quantum critical points and its corresponding susceptibility presents a peak.
The susceptibility peak is sharp and divergent at the N\'{e}el-Haldane critical point, which
grows almost linearly with $N$ while the corresponding peak at the Haldane-large-$D$ transition
is wide and grows sub-linearly with $N$. This is in agreement with the scaling argument
presented in Ref.~\onlinecite{CamposVenuti:2007}, which states that the quantum phase transition
does not always lead to a superextensive growth of fidelity susceptibility.
In contrast to Ref.~\onlinecite{Tzeng:2008} our exact diagonalization results show
similar qualitative behavior for all easy axis anisotropies, none of our results
fall into the scaling behavior $|D_m(N)-D_c| \sim N^{-1/\nu}$, which suggests two facts:
(i) The topological Haldane-large-$D$ phase transition has strong non-local quantum correlations
which requires an intensive numerical simulation \cite{Hu:2011} for a very large system size
to get the quantum critical point accurately.
(ii) The scaling behavior of fidelity could be different at small size limit $N \delta <1$
and large size one $N \delta >1$, similar to what has been observed for the Ising chain
in transverse field. \cite{Rams:2011,Langari:2012}

The exact diagonalization data of clusters with an odd number of sites ($N=9, 11, 13, 15$) unveils
a symmetry, which protects the topological Haldane-large-$D$ transition: The parity ``string order parameter'' in Fig.~\ref{Inversion-N-Odd}  shows an abrupt change of parity at the topological phase boundary.
This is a very accurate signature of the Haldane-large-$D$ phase transition on small chains, which
gives the quantum critical point accurately as obtained in Table.~\ref{t1}.
It suggests an investigation of a similar quantity to identify other types of topological phase transitions.
It has to be noticed that the expectation value of inversion operator ($\langle P_r \rangle$) is
a non-local quantity and thus expected be sensitive for a topological phase transition.
However, the expectation value of a single permutation operator like $\langle P_{8,9} \rangle$ (for $N=15$)
encounters a sign change at the phase boundary, which can be considered as a signature of the mentioned
phase transition. This type of order parameters is thus helpful to determine the phase boundary
of topological transition on small size clusters, which need less resources.

\begin{acknowledgments}
This work was supported in part by
the Office of Vice-President for Research of
Sharif University of Technology.
A. L. acknowledges the support from the Alexander von Humboldt Foundation.
\end{acknowledgments}

\appendix
\section{The QRG procedure \label{qrg-appendix}}

The renormalized Hamiltonian $H'$ (Eq.\ref{hprime}) is obtained within QRG
procedure using the embedding operator, Eq.\ref{embedding}, which is given by
\be
H'=\sum_{I=1}^{N/3} \big(T_I^{\dagger} h_I^B T_I +
T_I^{\dagger}T_{I+1}^{\dagger} h_{I, I+1}^{BB} T_{I+1}T_I \big).
\label{a1}
\ee
The first part of this projections leads to
\be
T_I^{\dagger} h_I^B T_I=E_0 \mathbf{1}+(E_1 - E_0) (S_I^z)^2,
\label{a2}
\ee
where $E_0$ and $E_1$ are the two lowest eigenvalues of the block Hamiltonian,
namely $E_0$ is the smallest root of the following equation,
\begin{equation}
E^3 + E^2 \left( \Delta -4 D \right) +
E \left(4 D^2 - 2\Delta D - 6 \right) + 8 D = 0,
\label{a3}
\end{equation}
and $E_1$ is the smallest solution of
\begin{eqnarray}
&& E^4 +
   E^3 \left( 2 \Delta - 8 D \right)
+ E^2 \left( 22 D^2 - 10 \Delta D - 5 \right) \nonumber \\
&& + E \left(-24 D^3 + 14 \Delta D^2 + 24 D - 6 \Delta \right) \nonumber \\
&& + 9 D^4 - 6 \Delta D^3 - 27 D^2 + 14 \Delta D = 0.
\label{a4}
\end{eqnarray}
The second term of Eq.\ref{a1} defines the effective interaction between blocks $I$ and $I+1$
in terms of renormalized operators
\bea
T_I^{\dagger} S_{I, j}^{\alpha} T_I&=& X_\text{ren} S_I^{'\alpha} \;;\; j=1, 3 \;;\; \alpha=x, y ,\nonumber \\
T_I^{\dagger} S_{I, j}^{z} T_I&=& Z_\text{ren} S_I^{'z} \;;\; j=1, 3,
\label{a5}
\eea
where $S_{I, j}^{\alpha}$ represents the $\alpha$ component of spin-1 at site-$j$ in the $I$-th block
of the original Hamiltonian
and $S_I^{'\alpha}$ is spin-1 operator defined for the $I$-th block in the renormalized Hilbert space.
The renormalization coefficients $X_\text{ren}$ and $Z_\text{ren}$ are given by the following expressions,
\begin{widetext}
\begin{eqnarray}
X_\text{ren} &=& \frac{1}{\sqrt{A_5 A_9}} \bigg[ 2 \left(E_0-2 D\right) + 2 \left(E_1-3 D\right) \left[4 D^2-2 D \left(\Delta +2 E_0\right)+E_0 \left(\Delta +E_0\right)-2\right] - \nonumber \\
&& A_2 A_3 \left(D - E_1\right) \big[4 D^2 E_0 - 2 A_8 - 2 D \left[E_0 \left(\Delta + 2 E_0\right) - 3\right] + E_0^2 \left(\Delta +E_0\right)-6 E_0 \big] + \nonumber \\
&& 2 A_1 A_2 A_3 \left(A_8 - D + E_0\right) - \frac{A_7 \left[A_1 A_2 A_3 \left(D - E_1\right) - 6 D + 2 E_1\right]}{2 D - E_0} \bigg], \nonumber \\
Z_\text{ren} &=& \frac{1}{A_5} \left[
   A_3^2 A_2^2 {\left(D - E_1\right)}^2 -
   {\big[\left(D - E_1\right) \left[A_1 A_2 A_3 \left(D - E_1\right) + 4 E_1 - 12 D\right] + 2 \big]}^2 +
4 {\left(E_1-3 D\right)}^2 + 4 \right].
\end{eqnarray}
\end{widetext}

\newpage

We have defined the following relations for th $A_i$ constants,
\begin{eqnarray}
A_1 &=& 2 \Delta - 3 D + E_1,  \nonumber \\
A_2 &=& E_1^2 - 4 D E_1 + 3 D^2 - 1, \nonumber \\
A_3 &=& \frac{1}{\Delta - 2 D + E_1}, \nonumber \\
A_4 &=& \frac{1}{\Delta - 2 D + E_0}, \nonumber \\
A_5 &=& A_3^2 {\left[ A_1 A_2 \left(D - E_1\right) + 2 \left(3 D - E_1\right) \left(2 D - \Delta - E_1\right) \right]}^2 \nonumber \\
    &&+ A_2^2 A_3^2 {\left(D - E_1\right)}^2 + A_1^2 A_2^2 A_3^2 + 4 {\left(E_1 - 3 D\right)}^2 + 4 + \nonumber \\
    && {\big[ \left(D - E_1\right) \left[A_1 A_2 A_3 \left(D - E_1\right) + 4 \left(E_1 - 3 D\right)\right] + 2 \big]}^2,  \nonumber \\
A_6 &=& 16 D^4 E_0 - 8 D^3 \left[2 E_0 \left(\Delta +2 E_0\right)-3\right] \nonumber \\
    &&+ 4 D^2 \left[6 E_0^3 + 6 \Delta E_0^2 + \left(\Delta ^2 - 12\right) E_0 - 3 \Delta \right] \nonumber \\
    &&- 2 D \left[ 4 E_0^4 + 6 \Delta E_0^3 + \left(2 \Delta^2 - 15\right) E_0^2 - 9 \Delta E_0 + 4\right] \nonumber \\
    &&+ E_0^2 \left(\Delta + E_0\right) \left[E_0 \left(\Delta + E_0\right) - 6\right] + 6 E_0,  \nonumber 
\end{eqnarray}
\begin{eqnarray}
A_7 &=& -E_0^3 + E_0^2 (4 D - \Delta ) + 2 E_0 (D \Delta -2 D^2) - 4 D + 2, \nonumber \\
A_8 &=& \frac{A_4 \left(3 E_0-4 D\right)}{2 D-E_0}, \nonumber\\
A_9 &=& \frac{A_4^2 A_6^2 + A_7^2}{{\left(E_0 - 2 D\right)}^2} + 2{\left(E_0 - 2 D\right)}^2 \nonumber \\
    &&+ 4{\left(A_8 - D + E_0\right)}^2 + 4 \nonumber \\
    &&+ {\left[4 D^2 - 2 D \left(\Delta + 2 E_0\right) + E_0 \left(\Delta + E_0\right) - 2\right]}^2.
\end{eqnarray}

Finally, the renormalized coupling constants, Eq.\ref{rg-flow} are given by the following relations,
\bea
J'&=& (X_\text{ren})^2 J , \nonumber \\
\Delta'&=&(\frac{Z_\text{ren}}{X_\text{ren}})^2 \Delta ,\nonumber \\
D'&=& \frac{E_1-E_0}{(X_\text{ren})^2}.
\eea
%

\bibliography{QRGS1XXZD}
\end{document}